\documentclass[12pt]{article}
\usepackage{amssymb,graphicx}
\usepackage{epstopdf}
\usepackage{amsmath,amsfonts}
\usepackage{epsfig}

\def\e{{\rm e}}

\def\d{\partial}
\def\l{\left(}
\def\r{\right)}

\newcommand{\be}{\begin{equation}}
\newcommand{\ee}{\end{equation}}
\newcommand{\bea}{\begin{eqnarray}}
\newcommand{\eea}{\end{eqnarray}}
\newcommand{\bg}{\begin{gather}}
\newcommand{\eg}{\end{gather}}
\newcommand{\bseq}{\begin{subequations}}
\newcommand{\eseq}{\end{subequations}}

\renewcommand{\ln}{\mathop{\rm ln}\nolimits}

\begin{document}
\begin{flushright}
\end{flushright}
\vspace{10pt}
\begin{center}
  {\LARGE \bf Consistent NEC-violation: \\[0.3cm] towards 
creating a universe\\ [0.3cm] in the
laboratory } \\
\vspace{20pt}
V. A. Rubakov$^{a,b}$

\vspace{15pt}
$^a$\textit{
Institute for Nuclear Research of
         the Russian Academy of Sciences,\\  60th October Anniversary
  Prospect, 7a, 117312 Moscow, Russia}\\
\vspace{5pt}
$^b$\textit{
Department of Particle Physics and Cosmology,
Physics Faculty, \\
M.V. Lomonosov Moscow State University,\\
Vorobjevy Gory, 119991, Moscow, Russia
}

    \end{center}
    \vspace{5pt}

\begin{abstract}
Null Energy Condition (NEC) can be violated in a consistent
way in models with unconventional kinetic terms, notably, in
Galileon theories and their generalizations. We make use of
one of these,
the scale-invariant
kinetic braiding model, to discuss whether a universe
can in principle
be created by man-made processes. We find that even though
the simplest models of this sort can have both healthy
Minkowski vacuum and consistent NEC-violating phase,
there is an obstruction for creating a universe in a straightforward
fashion. To get around this obstruction, we design a more complicated model,
and present a scenario for the creation of a universe in the
laboratory.
\end{abstract}

\section{Introduction}
\label{sec:intro}

Once it was realized that inflation can stretch
a tiny region of space into the entire visible Universe,
a question has been naturally raised of whether one can in principle 
create a new universe by man-made processes~\cite{Berezin:1984vy,Farhi:1986ty}.
In the context of classical General Relativity and conventional
theories of matter obeying the Null Energy Condition (NEC),
the answer is negative~\cite{Farhi:1986ty,Berezin:1987ep} 
because of the problem with the initial singularity
guaranteed by the Penrose theorem~\cite{Penrose} (see, however,
Refs.~\cite{Borde:1998wa,Sakai:2006fg}). Widely discussed ways out
are to invoke 
tunneling~\cite{Berezin:1987ea,Farhi:1989yr,Fischler:1989se,Linde:1991sk,Garriga:1997ef,Garriga:1999hf,Dutta:2005gt,Aguirre:2005nt,Lee:2006vka,Piao:2007cj}
or other quantum 
effects~\cite{Frolov:1988vj,Guendelman:2010pr,Lukash:2013ts},
modify gravity~\cite{Mukhanov:1991zn,Brandenberger:1993ef,Trodden:1993dm}
and violate NEC~\cite{Lee:2007dh,Guendelman:2008ys,Yeom:2009mn,Hwang:2010gc}.
The latter option, however, has been problematic, since most of
the NEC-violating theories are plagued by pathologies like ghosts,
gradient instability and/or superluminality. 
Yet it has been realized some time ago~\cite{Senatore:2004rj,VR-vector,Creminelli:2006xe,Nicolis:2009qm,Genesis1,Genesis2} that within General
Relativity,
the NEC-violation
is not necessarily accompanied by unacceptable 
pathologies,
if one considers theories with unconventional
kinetic terms. One class of examples is given by 
the
Galileon theory~\cite{Nicolis:2008in} and its 
generalizations~\cite{deRham:2010eu,Deffayet:2010qz,Kobayashi:2010cm,Goon-prl,Goon:2011qf,Deffayet:2010zh,Kamada:2010qe,Pujolas:2011he}.
Indeed, by making use of the Galileon,
a cosmological Genesis model has been constructed~\cite{Genesis1,Genesis2},
in which the evolution
starts from nearly Minkowski space-time,
the energy density eventually  builds up and the Universe enters the
epoch of rapid expansion. The NEC-violation in this scenario
occurs in a controllable and consistent way~\cite{Genesis2}.

These developments suggest that one might be able to create a universe
in the laboratory in a purely classical way and within General Relativity.
In this paper we suggest a scenario of this sort, allowing ourselves
not only to set up appropriate initial conditions for the field evolution,
but also 
to design a field theoretic model at our will. The idea is to
construct initial condition in a Galileon-type theory such that inside
a large sphere the field is nearly homogeneous and behaves like
at the initial stage of
Genesis, whereas outside this sphere the field tends to a
constant and space-time is asymptotically Minkowskian.
For this initial data, 
the energy density and pressure are initially small everywhere and
the entire space-time is nearly Minkowskian, so that the required
field configuration can in principle be prepared in the laboratory. 
As the field evolves from this initial state according to
its equation of motion, the energy density inside the large sphere
increases, space undergoes accelerated expansion there, and the region 
inside the sphere eventually 
becomes a man-made universe. Outside this sphere the energy density
remains small and asymptotes to zero at large distances;
the space-time is always asymptotically Minkowskian.

Implementing this idea is not entirely trivial, however.
The field theoretic model we are after should have not only 
healthy Genesis regime but also healthy Minkowski vacuum.
The latter property is lacking in the model of Refs.~\cite{Genesis1,Genesis2}.
Moreover, there must be smooth and healthy interpolation
between the Genesis regime inside the large sphere
and asymptotic Minkowski vacuum; we will see that this requirement
is particularly restrictive. For this reason the model we end up
with is rather contrived. Yet it serves the purpose of proof-of-principle.

This paper is organized as follows. We find it instructive to begin 
in Sec.~\ref{2-all}
with
a prototype model which actually does not work. We introduce the model 
and collect useful formulas in 
Sec.~\ref{2-prel}, consider the stability of the Minkowski vacuum
in Sec.~\ref{2-Mink} and study a NEC-violating homogeneous 
solution in Minkowski space-time in Sec.~\ref{2-rolling}.
We find in Sec.~\ref{sec:obstruction} that creating 
a universe in the laboratory in a way outlined above is actually
impossible in the model we consider in Sec.~\ref{2-all} and, in fact,
the obstruction we encounter is inherent in a class of
NEC-violating theories. Yet we are able to design a working
model in Sec.~\ref{improved-all} by introducing an extra field
whose background produces spatially inhomogeneous couplings.
We present the model and discuss
relevant stability issues in Sec.~\ref{sec:improved1}, and 
end up with 
a fairly detailed scenario for the creation of a universe in the
laboratory in Sec.~\ref{improved-scenario}. We conclude in
Sec.~\ref{concl}.

\section{Prototype model}
\label{2-all}

\subsection{Preliminaries}
\label{2-prel}

In this Section we consider a model 
of kinetic braiding 
type~\cite{Deffayet:2010qz,Kobayashi:2010cm} with a scalar field $\pi$,
and impose
dilatation invariance of the action in Minkowski space-time,
\be
\pi (x) \to \pi^\prime(x) = \pi(\lambda x) + \ln \lambda \; .
\label{may4-13-5}
\ee 
This invariance, albeit {\it ad hoc}, simplifies the analysis
considerably. 
The dilatationally-invariant kinetic brading Lagrangian 
is (mostly negative signature)
\be
L = F (Y) \e^{4\pi} + K(Y) \Box \pi \cdot \e^{2\pi} \; ,
\label{may3-13-1}
\ee
where
\be
Y = \e^{-2\pi} (\d \pi)^2
\label{mar5-13-2}
\ee
and $F$ and $K$ are yet unspecified functions.
Assuming that $K$ is analytic near the origin, we  set 
\be
K(Y=0)= 0
\label{may6-13-1} \; .
\ee 
Indeed, upon integrating by parts, a constant part of $K$
can be 
absorbed into the $F$-term in
\eqref{may3-13-1}. 

The field equation is
\begin{align}
& 4 \e^{4\pi} F - 2 \e^{2\pi} (\d \pi)^2 F^\prime -
2 \d_\mu \l \e^{2\pi} F^\prime \d_\mu \pi \r 
\\
& + 2 \e^{2\pi} \Box \pi \cdot K  
+ \Box \l \e^{2\pi} K \r - 
2 \Box \pi \cdot (\d \pi)^2 K^\prime
- 2 \d_\mu \l \Box \pi \cdot K^\prime \d_\mu \pi \r = 0 \; ,
\label{mar5-13-1}
\end{align}
where prime denotes $d/dY$. 
Let $\pi_c (x)$ be a solution to this equation.
We will be interested also in perturbations about $\pi_c$.
To this end, let us decompose  $\pi = \pi_c + \chi$ and write
the quadratic Lagrangian for perturbations:
\begin{align}
L^{(2)} &= \l \d \chi \r^2 F^\prime \e^{2\pi_c} + 2 
F^{\prime \prime}
 \d_\mu \pi_c \d_\nu \pi_c \cdot \d_\mu \chi \d_\nu \chi
\nonumber \\
& + \l \d \chi \r^2 \left[-2 K \e^{2 \pi_c} + 2 \l \d \pi_c \r^2 K^\prime
+ \d_\mu \l K^\prime \d_\mu \pi_c \r + \Box \pi_c \cdot K^\prime \right]
\nonumber \\
& 
+ \d_\mu \chi \d_\nu \chi \left[
-2 \d_\nu \l K^\prime \d_\mu \pi_c \r + 2 \Box \pi_c \e^{-2\pi_c}
 K^{\prime \prime} \d_\mu \pi_c \d_\nu \pi_c \right]
\nonumber\\
&+ \chi^2 \left[ 8 F \e^{4\pi_c} - 6 F^\prime \e^{2\pi_c} (\d \pi_c)^2
- 2 \d_\mu \l F^\prime \e^{2\pi_c} \d_\mu \pi_c \r
+ 2 F^{\prime \prime} (\d \pi_c)^4 + 2 \d_\mu \l F^{\prime \prime}
(\d \pi_c)^2 \d_\mu \pi_c \r \right]
\nonumber\\
& + \chi^2 \left[
\Box \l \e^{2\pi_c} K \r + 2K \e^{2\pi_c}\Box \pi_c
- \Box \l K^\prime (\d \pi_c)^2 \r
-2 \Box \pi_c (\d \pi_c)^2 K^\prime \right.
\nonumber\\
& \left. + 2 \Box \pi_c \e^{-2\pi_c} (\d \pi_c)^4 K^{\prime \prime}
+ 2 \d_\mu \l \Box \pi_c \e^{-2\pi_c} (\d \pi_c)^2 K^{\prime \prime}
\d_\mu \pi_c \r \right]  \; .
\label{feb27-13-2}
\end{align}

We will eventually need the expression for the energy-momentum tensor.
To this end, we consider minimal coupling to the metric, i.e., set
$Y = \e^{-2\pi} g^{\mu \nu} \d_\mu \pi \d_\nu \pi$ and $\Box \pi =
\nabla^\mu \nabla_\mu \pi$ in curved space-time. To calculate the 
energy-momentum 
tensor, we note that in curved space-time,
the $K$-term in $\sqrt{-g} L$ 
can be written, upon integrating by parts, as
$\sqrt{-g} g^{\mu \nu} \d_\mu \pi \d_\nu \l K \e^{2\pi} \r$.
Then the
variation with respect to $g^{\mu \nu}$ is straightforward, and we get
\begin{align*}
T_{\mu \nu} &= 2 F^\prime \e^{2\pi} \d_\mu \pi \d_\nu \pi - g_{\mu \nu} F \e^{4\pi}
\nonumber
\\& + 2 \Box \pi \cdot K^\prime  \d_\mu \pi \d_\nu \pi -
\d_\mu \pi \cdot \d_\nu \l K \e^{2\pi} \r-
\d_\nu \pi \cdot \d_\mu \l K \e^{2\pi} \r
+ g_{\mu \nu} g^{\lambda \rho} \d_\lambda \pi \d_\rho \l K \e^{2\pi} \r \; .
\end{align*}

In what follows we mostly consider homogeneous backgrounds, $\pi = \pi (t)$,
and omit subscript $c$ wherever possible. For a homogeneous field,
equation of motion~\eqref{mar5-13-1} reads
\begin{align}
& 4 \e^{4\pi} F + F^\prime \e^{2\pi}\left(- 6 \dot{\pi}^2
- 2\ddot{\pi} \right) - 
 2 \e^{2\pi} \dot{\pi}  F^{\prime \prime} \dot{Y}
\nonumber\\
 & + K \e^{2\pi} \left( 4 \dot{ \pi}^2 + 4 \ddot{\pi} \right)
+ 4 \e^{2\pi} \dot{ \pi} K^\prime \dot{Y}
 + K^{\prime \prime} \dot{Y} \left( -2  \dot{ \pi}^3 \right)
 + K^\prime \left( - 12\dot{ \pi}^2  \ddot{\pi} + 4 \dot{ \pi}^4
\right)
=0 \; ,
\label{mar5-13-3}
\end{align}
while the energy density and pressure are
\begin{subequations}
\begin{align}
\rho &= \e^{4\pi} Z
\label{may3-13-11}
\\
p &= \e^{4\pi} \l F - 2YK - \e^{-2\pi} K^\prime \dot{\pi} \dot{Y} \r \; ,
\label{may5-13-14}
\end{align}
\end{subequations}
where
\[
Z= -F + 2Y F^\prime - 2YK + 2Y^2K^\prime \; .
\]
It is straightforward to see that for 
$\dot{\pi} \neq 0$, eq.~\eqref{mar5-13-3}
is equivalent to energy conservation, $\dot{\rho} = 0$.
Finally, in a homogeneous background the quadratic Lagrangian for 
perturbations, eq.~\eqref{feb27-13-2}, simplifies to
\be
L^{(2)} = U \dot{\chi}^2 - V (\d_i\chi)^2 + W \chi^2 \; ,
\label{may3-13-10}
\ee
where
\begin{subequations}
\begin{align}
U &= \mbox{e}^{2\pi_c} \l F^\prime + 2 Y F^{\prime \prime}
- 2K + 2 Y K^\prime + 2 Y^2 K^{\prime \prime} \r
= \e^{2\pi_c} Z^\prime \; ,
\label{may4-13-2}
\\
V &= \e^{2\pi_c}\l F^\prime - 2K + 2YK^\prime -
2 Y^2 K^{\prime \prime} \r + \l 2K^\prime + 2Y K^{\prime \prime}
\r \ddot{\pi}_c \; .
\end{align}
\label{may3-13-2}
\end{subequations}
We will not need the general expression for $W$.

\subsection{Minkowski vacuum}
\label{2-Mink}

Recalling that $K(0)=0$, we find that the Minkowski vacuum
$\d \pi = 0$ exists (cosmological constant is zero), provided that
\be
     F(0) = 0 \; .
\label{may3-13-6}
\ee
It is clear from eq.~\eqref{may3-13-2} that it
is stable for
\be
F^\prime (0) > 0 \; .
\label{may3-13-7}
\ee
There remains an issue~\cite{Adams:2006sv,Nicolis:2009qm}
 of the possible superluminality of
perturbations about
backgrounds in the neighbourhood of the Minkowski 
vacuum,
i.e., backgrounds
with small $\d \pi_c (x)$. 
From this viewpoint,
the most dangerous terms in \eqref{feb27-13-2} 
involve $\d_\mu \chi \d_\nu \chi \, \d_\mu \d_\nu \pi_c$.
We make these terms small by requiring that
\be
K^\prime (0) = 0 \; .
\label{may3-13-8}
\ee
Then the inverse effective metric for perturbations, 
modulo irrelevant terms, is
\[
G^{\mu \nu} = \eta^{\mu \nu} +
\frac{1}{F^\prime \e^{2\pi_c}}
\left[  2 
F^{\prime \prime}
 \d^\mu \pi_c \d^\nu \pi_c 
-\d^\nu \l K^\prime \d^\mu \pi_c \r  - \d^\mu \l K^\prime \d^\nu \pi_c \r \right]
\; .
\]
and the metric itself reads
\be
G_{\mu \nu} = \eta_{\mu \nu} -
\frac{1}{F^\prime \e^{2\pi_c}}
\left[  2 
F^{\prime \prime}
 \d_\mu \pi_c \d_\nu \pi_c 
-\d_\nu \l K^\prime \d_\mu \pi_c \r  - \d_\mu \l K^\prime \d_\nu \pi_c \r
\right]
\label{may5-13-10}
\ee
Potentially dangerous situation is when  null  (in conventional sense)
direction of propagation
$n^\mu$  is time-like in the metric
$G_{\mu \nu}$. For generic $n^\mu$ this is avoided
by requiring
\be
F^{\prime \prime}(0) > 0 \; .
\label{may3-13-9}
\ee
Indeed, 
the dangerous terms are of order $K^{\prime \prime}(\d \pi_c)^2 \d^2 \pi_c$, so
the first term in square brackets is the dominant source of Lorentz-violation
and $G_{\mu \nu}n^\mu n^\nu < 0$ for generic  $n^\mu$.

This argument does not apply to the special direction for which
$\d_\mu \pi_c \, n^\mu= 0$. 
Let us consider this direction separately.
We treat our model
near the Minkowski vacuum as a low energy effective theory
with a UV cutoff $\Lambda$. 
Consider now the background configuration (we set $\pi_c (x=0) = 0$
by using the dilatation symmetry)
\[
\pi_c = q_\mu x^\mu + \frac{1}{2} A_{\mu \nu} x^\mu x^\nu
\]
and choose the wave vector $k^\mu = n^\mu k$ 
such that $q_\mu k^\mu = 0$. Then 
the effective metric~\eqref{may5-13-10}
at distance $l$ from the origin in the direction $n^\mu$
is
\be
G_{\mu \nu} = \eta_{\mu \nu} -
\frac{1}{F^\prime(0)} 
\l  2 
F^{\prime \prime} (0) l^2
 A_{\mu \lambda} n^\lambda A_{\nu \rho} n^\rho 
-2 K^{\prime \prime}(0)   q^2 A_{\mu \nu}  
\r \; .
\label{mar1-13-1}
\ee
We see that $G_{\mu \nu} n^\mu n^\nu > 0$
 near the origin,
if $A_{\mu \nu} n^\mu n^\nu \equiv (n\cdot A \cdot n) < 0$ 
(assuming for definiteness that
$K^{\prime \prime} (0)> 0$ and $q^2<0$), which signalizes the
superluminality. 
Near the origin the
correcion to the propagation speed 
is of order
\[
\delta c \sim \frac{K^{\prime \prime} (0)}{F^\prime(0)} q^2 (n \cdot A\cdot n)
\; .
\]
This correction becomes detectable when it yields the
deviation from 
distance traveled by light which is at least of the order of
the wavelength~\cite{Nicolis:2009qm},
\[
\delta c \cdot l \gtrsim k^{-1} \; .
\]
We require that at this distance the first term in parenthesis in
\eqref{mar1-13-1}, which {\it reduces} the speed of signal, 
dominates, 
\[
F^{\prime \prime} (0) l^2
 (n \cdot A \cdot n)^2 \sim \frac{F^{\prime \prime}(0) F^{\prime \, 2}(0)}{k^2
K^{\prime \prime \, 2}(0) q^4} 
> K^{\prime \prime}(0) q^2 (n\cdot A\cdot n) \; .
\]
For $k^2, q^2, A_{\mu \nu} \ll  \Lambda^2$,
 this inequality 
holds, provided that the functions $F$ and $K$ obey a
constraint
\be
\frac{F^{\prime \prime}(0) F^{\prime \, 2}(0)}{K^{\prime \prime \, 3}(0)}
\gtrsim \Lambda^{10} \; .
\label{may5-13-20}
\ee
Under this constraint,
the local superluminality
is undetectable, and hence not dangerous.

We conclude that the Minkowski vacuum and its neighborhood are healthy,
provided that eqs.~\eqref{may3-13-6}, \eqref{may3-13-7},  \eqref{may3-13-8},
\eqref{may3-13-9} and \eqref{may5-13-20}
are satisfied.

\subsection{Rolling solution}
\label{2-rolling}

With an appropriate choice of the functions $F$ and $K$,
eq.~\eqref{mar5-13-3} admits also a rolling solution, similar to
that in the Galileon theory~\cite{Nicolis:2009qm,Genesis1,Genesis2}:
\be
\e^\pi = \frac{1}{\sqrt{Y_*}(t_* - t)} \; ,
\label{may4-13-1}
\ee
where $t_*$ is an arbitrary constant.
For this solution
$Y =Y_* = \mbox{const}$,
and $Y_{*}$ is determined from equation 
\be
Z(Y_*) \equiv -F + 2Y_{*} F^\prime - 2 Y_{*}  K + 2 Y^2_{*} K^\prime  = 0 \; ,
\label{feb27-13-1}
\ee
where $F$, $F^\prime$, etc., are evaluated at $Y=Y_*$.
For this solution one has $T_{00}= \rho =0$ and
\be
p = \frac{1}{Y_*^2(t_* - t)^4} \l F - 2Y_* K \r \; .
\label{mar3-13-2}
\ee
Thus, the rolling background violates NEC, provided that
\be
\mbox{NEC violation:}~~~~~~~~~~~~~~~~
2 Y_{*}K - F >0 \; .
\label{mar3-13-6}
\ee

The quadratic Lagrangian
for perturbations \eqref{may3-13-10} reduces in this background to
\be
L^{(2)} = \frac{A}{Y_{*}(t_* - t)^2} [\dot{\chi}^2 -(\d_i \chi)^2] + 
 \frac{B}{Y_{*}(t_* - t)^2} \dot{\chi}^2 +  
\frac{C}{Y_{*}^2(t_* - t)^4} \chi^2 \; ,
\label{mar3-12-1}
\ee
where 
\begin{align*}
A&=\e^{-2\pi_c} V = F^\prime - 2 K + 4 Y_{*} K^\prime 
\\
B &= \e^{-2\pi_c}(U-V)=
 2 Y_{*}F^{\prime \prime} - 2 Y_{*} K^\prime + 2 Y^2_{*} K^{\prime \prime}
\\
C &= 8 F - 12 Y_{*} F^\prime + 8 Y_{*}^2 F^{\prime \prime}
+ 8 Y_{*} K - 8 Y_{*}^2 K^\prime + 8 Y^3_{*} K^{\prime \prime}
\end{align*}
are time-independent coefficients. 
As a cross check, one can derive from the latter Lagrangian
the equation for homogeneous perturbation
$\chi(t)$ about the rolling background and see that $\chi = \d_{t} \pi_c =
(t_* - t)^{-1}$ obeys this equation, as it should. Indeed, making use of
eq.~\eqref{feb27-13-1} one finds that the
coefficients of $\dot{\chi}^2$
and $\chi^2$ in eq.~\eqref{mar3-12-1} are related in a simple way,
\[
4(A+B) = C/Y_{*} \; .
\]
Hence, homogeneous perturbation obeys a universal equation
\[
- \frac{d}{dt} \l \frac{\dot{\chi}}{(t_* - t)^2} \r
+ 
 4\frac{\chi}{(t_* - t)^4} = 0 \; ,
\]
whose solutions are $\chi = (t_* - t)^{-1}$ and $\chi= (t_* - t)^4$.
This shows that the rolling background is stable against low momentum 
perturbations; like in the Galileon case~\cite{Genesis1},
the growing perturbation $\chi = (t_* - t)^{-1}
\cdot \chi_0({\bf x})$ with slowly varying  $\chi_0({\bf x})$
can be absorbed into slightly inhomogeneous time shift.

In fact, we can see in more general terms that the rolling background
is an attractor in the class of homogeneous solutions. To this end, we
use the conservation of energy \eqref{may3-13-11} 
to write for any homogeneous solution
\be
\e^{4\pi} Z =  C  
=\mbox{const}  \; .
\label{apr19-13-2}
\ee
Now, the relation $\dot{\rho} = 0$ for positive
$\dot{\pi}$ can be written
as
\[
4 \dot{\pi} Z + \dot{Y} Z^\prime = 4 \e^\pi Y^{1/2} Z  + \dot{Y} Z^\prime
= 4 \l \frac{|C|}{|Z|} \r^{1/4}  Y^{1/2} Z  + \dot{Y} Z^\prime = 0 \; .
\]
If $Z^\prime \neq 0$, this gives
\be
\dot{Y} = - 4 \l \frac{|C|}{|Z|} \r^{1/4} Y^{1/2} \frac{Z}{Z^\prime} \; ,
\label{mar5-13-9}
\ee
so that
\be
\dot{Z} = - 4 |C|^{1/4} Y^{1/2} \frac{Z}{|Z|^{1/4}} \; .
\label{may5-13-3}
\ee
This shows that the rolling 
solution with $Z=0$ and $\dot{\pi} > 0$ is an attractor
whose basin of attraction is bounded by the points, if any,
where $Z^\prime (Y) = 0$.
This is also obvious from eq.~\eqref{apr19-13-2}:
if $\pi$ increases, $|Z|$ decreases.

Let us consider the stability of the rolling background and
subluminality of the perturbations about it. 
The spatial gradient term  in \eqref{mar3-12-1}
has correct (negative) sign provided that
\be
\mbox{No~gradient~instability:}~~~~~~~~~~~~~~~~
 A=F^\prime - 2 K + 4 Y_{*} K^\prime > 0 \; .
\label{mar3-13-4}
\ee
The speed of perturbations about the rolling background is smaller
than the speed of light, if the coefficient of $\dot{\chi}^2$
is greater than that of $-(\d_i \chi)^2$, i.e.,
\be
\mbox{Subluminality:}~~~~~~~~~~~~~~~~
B=2Y_* F^{\prime \prime} - 2Y_* K^\prime + 2Y^2_{*} K^{\prime \prime} > 0 \; .
\label{mar3-13-5}
\ee
We require that the latter inequality holds in a strong sense,
then the perturbations about the rolling solution are strictly
subluminal, and hence the perturbations about backgrounds
neighbouring the rolling solution
are subluminal as well. 
When both inequalities \eqref{mar3-13-4} and  \eqref{mar3-13-5}
are satisfied,
there are no ghosts either.
The conditions \eqref{mar3-13-6}, \eqref{mar3-13-4} and 
\eqref{mar3-13-5} together with eq.~\eqref{feb27-13-1} can be satisfied
at $Y=Y_*$
by a judicious choice of the functions $F$ and $K$ in the neighbourhood
of this point, so that the NEC-violation is stable and subluminal.
This can be seen as follows. Equation~\eqref{feb27-13-1}
can be used to express $F(Y_*)$ in terms of $F^\prime (Y_*)$, $K(Y_*)$ and
$K^\prime (Y_*)$, namely, $F=2Y_* F^\prime - 2 Y_* K + 2 Y_*^2 K^\prime$.
Then the inequalities  \eqref{mar3-13-6}, \eqref{mar3-13-4} are satisfied,
provided that $2 K - 4Y_* K^\prime < F^\prime < 2 K - Y_* K^\prime$, 
which is possible
for positive $K^\prime$. The condition \eqref{mar3-13-5} can be satisfied
by an appropriate choice of $F^{\prime \prime}$ and $K^{\prime \prime}$.

Obviously, the functions $F(Y)$, $K(Y)$ 
can be chosen
in such a way that both Minkowski vacuum and rolling solution are
stable and healthy\footnote{This does not mean, though, that the entire model is
completely healthy: it can be a low energy effective theory of some
Lorentz-invariant UV-complete theory only if perturbations about {\it any}
allowed background are subluminal~\cite{Adams:2006sv}. This property
should hold also in the presence of gravity, cf. Ref.~\cite{Easson:2013bda}.
The analysis of this issue is beyond the scope of this paper.}, i.e., 
eqs.~\eqref{may3-13-6}, \eqref{may3-13-7},  \eqref{may3-13-8},
\eqref{may3-13-9} and \eqref{may5-13-20}
are satisfied at $Y=0$ and 
eqs.~\eqref{feb27-13-1}, \eqref{mar3-13-6}, \eqref{mar3-13-4} and 
\eqref{mar3-13-5} are satisfied at $Y=Y_*$.  With such a choice
of $F(Y)$, $K(Y)$,
both 
Minkowski vacuum and rolling solution are attractors, with 
non-overlapping basins of attraction. 

\subsection{Obstruction to a simple way of creating a
universe in the laboratory}
\label{sec:obstruction}

It is now tempting to implement the approach outlined in
Sec.~\ref{sec:intro} in a simple way, by
considering the initial field
$\pi (t, {\bf x})$ which slowly varies
in space and interpolates between the rolling solution 
\eqref{may4-13-1}
inside a large
sphere and Minkowski vacuum $\d \pi = 0$ at spatial infinity.
By slow variation in space we mean that the spatial derivatives of
$\pi$ are negligible compared to temporal ones, so that at each 
point in space $\pi$ evolves in the same way as in the homogeneous
case.

An advantage of this quasi-homogeneous approach is simplicity
of the analysis; a disadvantage is that it actually does not work in our 
prototype model. 
The point is that irrespectively of the equation of motion,
the term with $\dot{\chi}^2$ in \eqref{may3-13-10} 
is proportional to $Z^\prime (Y)$, see
eq.~\eqref{may4-13-2}.
Thus, for configurations slowly varying in space,
the absence of ghosts requires
\[
Z^\prime (Y) > 0
\]
everywhere. For both Minkowski vacuum and rolling solution we have $Z=0$,
so there is no ghost-free configuration that slowly
varies in space and interpolates between the two solutions
as $r$,
the distance from the center of the sphere, changes from
0 to $\infty$. 

This obstruction to have a quasi-homogeneous ghost-free configuration,
interpolating between
two different solutions of zero energy,
is generic in theories with the following
properties: (i) there is a single scalar field $\pi$;
(ii) the field equation is second order;
(iii) the action is invariant under dilatations \eqref{may4-13-5}.
This class of theories includes, e.g., conformal higher order
Galileons
of Ref.~\cite{Nicolis:2008in} and conformal 
DBI Galileons of Ref.~\cite{deRham:2010eu}.  
The argument is essentially the same as above. The Noether energy-momentum
tensor obeys
\[
\d_\mu T^{\mu}_{\nu} = - (\mbox{E.O.M.})\cdot \d_\nu \pi \; ,
\]
where  $(\mbox{E.O.M.})$ stands for the equation of motion.
Therefore, 
the equation of motion
for spatially homogeneous $\pi = \pi (t)$
is 
\be
(\mbox{E.O.M.}) = - \frac{1}{\dot{\pi}} \dot{\rho} \; .
\label{may5-13-1}
\ee
Since the field equation is second order, $\rho = \rho(\pi, \dot{\pi})$ 
does not contain $\ddot{\pi}$ and higher derivatives, and by
scale invariance it has the form $\rho = \exp(4\pi) \, Z(Y)$, where
$Y= \dot{\pi}^2 \, \exp(-2\pi) $, cf. eq.~\eqref{mar5-13-2},
and $Z$ is a model-dependent
function. It follows from eq.~\eqref{may5-13-1} that
the equation of motion for homogeneous perturbation about
the background $\pi_c (t)$ reads
\[
 - \frac{1}{\dot{\pi}_c} \frac{\d \rho}{\d \dot{\pi}_c} \ddot{\chi} + \dots 
= 0 \; ,
\]
where omitted terms do not contain $\ddot{\chi}$. Hence, the kinetic part
of the Lagrangian
for the perturbations has the form
\[
L^{(2)} \supset
\frac{1}{2\dot{\pi}_c}  \frac{\d \rho}{\d \dot{\pi}_c} \dot{\chi}^2
= \e^{2\pi_c} Z^\prime (Y)  \dot{\chi}^2 \; ,
\]
which is the same as in \eqref{may3-13-10}.
Both zero energy solutions have $Z=0$, so 
an interpolating configuration has $Z^\prime < 0$ somewhere
in between, and thus it is not ghost-free.

One way to get around this obstacle would be to insist on
slow spatial variation of the initial field configuration but
give up the prescription that the field inside the large sphere
is in the rolling regime \eqref{may4-13-1}. Instead, one would consider
the field with non-zero energy density inside the sphere,
so that there exists a smooth and ghost-free configuration
that interpolates, as $r$ increases, between this field and the
asymptotic Minkowski vacuum. This can hardly lead to the creation
of a universe, however, since eq.~\eqref{may5-13-3} shows that
the point $Y = 0$ is an attractor, and the field
in the interior of the sphere will likely relax to it\footnote{A
loophole here is that we neglect effects of gravity.}.
Near $Y=0$ one has $F = F^\prime (0) Y$ and $Z = F^\prime (0) Y$,
so that the equation of state is $p \approx \rho$ (recall that $K(0) =0$;
one can show that the last term in \eqref{may5-13-14} is negligible
at small $Y$). Thus, NEC does not get violated.

Other possibilities are to consider field configurations with
non-negligible spatial gradients or give up scale-invariance of
the action. In both cases
the above no-go 
agrument would be
irrelevant, but the analysis would be more complicated. We will follow
another route, and complicate the model instead.

\section{Improved model}
\label{improved-all}

\subsection{Spatially inhomogeneous couplings}
\label{sec:improved1}

We do not abandon quasi-homogeneity, but now  
allow the functions $F$ and $K$ to depend explicitly on spatial
coordinates. This can be the case 
if there is another field, call it $\varphi$, which 
determines the couplings entering these functions, and this field
acts as quasi-homogeneous background, $\varphi = \varphi({\bf x})$.
In this case one can consider a field configuration
$\pi (t, {\bf x})$ which at any point in space is approximately
given by the rolling solution \eqref{may4-13-1}, but with $Y_*$
depending on ${\bf x}$  (recall that $Y_*$ is independent
of time for the homogeneous solution). 
We prepare the background $\varphi ({\bf x})$ in such a way
that $Y_*({\bf x})$ is constant inside the large sphere
(to evolve into a man-made universe) and gradually aproaches zero
as $r \to \infty$. We have to check that with an appropriate
choice of the functions $F(Y;\varphi)$, $K(Y;\varphi)$, this 
construction is healthy everywhere in space, i.e.,
there are no pathologies inside the large sphere, at spatial infinity
and in the intermediate region (``the wall'').

Let $\Phi (\varphi)$ be a function of the new field, such that
$Y_* = \Phi(\varphi)$ is a solution to eq.~\eqref{feb27-13-1}.
As $r$ varies from zero to infinity, $\Phi({\bf x})$ changes from
some positive value $\Phi_0$ to zero. We are going to check that
the inequalities \eqref{mar3-13-4},  \eqref{mar3-13-5} can be satisfied
for any $\Phi \in (0, \Phi_0)$, so that there is no ghost or gradient
instability  anywhere in space (including the wall region), 
and propagation of perturbations
is subluminal, also in any region of  space.
To this end,
we write $F$ and $K$ in the vicinity
of $Y_*=\Phi$ as a series in $(Y-\Phi)$:
\begin{subequations}
\begin{align}
F &= a(\Phi) + b(\Phi) (Y - \Phi) + \frac{c(\Phi)}{2} (Y - \Phi)^2
\label{apr19-13-6a}
\\
K &= \kappa (\Phi) + \beta (\Phi) (Y-\Phi) + \frac{\gamma (\Phi)}{2}
(Y - \Phi)^2 \; ,
\label{apr19-13-6b}
\end{align}
\label{may5-13-11}
\end{subequations}
and set $\kappa (0) = 0$ without loss of generality, see 
eq.~\eqref{may6-13-1}.
In these terms,  eq.~\eqref{feb27-13-1}
reads
\be
a - 2\Phi b + 2\Phi \kappa - 2 \Phi^2 \beta = 0 \; ,
\label{apr19-13-10}
\ee
and the inequalities  \eqref{mar3-13-4},  \eqref{mar3-13-5} become
\begin{subequations}
\begin{align}
\mbox{No~gradient~instability:}~~~~~~~~~~~~~~~
 &  b(\Phi) - 2\kappa (\Phi) + 4 \Phi \beta (\Phi) > 0 \; 
\label{may5-13-12}
\\
\mbox{Subluminality:}~~~~~~~~~~~~~~~
 &  c (\Phi) - \beta (\Phi) +  \Phi \gamma (\Phi) > 0 \; .
\end{align}
\end{subequations}
Let us also write the pressure \eqref{mar3-13-2}
for the rolling solution:
\be
 p = 
\frac{1}{\Phi^2 (t_* - t)^4} (a - 2\kappa \Phi)
= \frac{1}{\Phi^2 (t_* - t)^4} \cdot 2\Phi (b - 2\kappa + \beta \Phi) \; ,
\label{apr19-13-7}
\ee
where we used eq.~\eqref{apr19-13-10}.
We require the NEC-violation inside the large sphere, i.e.,
\be
\mbox{NEC violation:}~~~~~~~~~~~~~~~~
b (\Phi_0) - 2\kappa (\Phi_0) +  \Phi_0 \beta(\Phi_0) < 0 \; .
\label{may5-13-13}
\ee
Finally, the stability conditions
 of the 
Minkowski vacuum, eqs.~\eqref{may3-13-7}, \eqref{may3-13-8},
read
\[
  b(0) >0 \; , \;\;\;\;\;\; \beta (0) = 0 \; , 
\]
while eq.~\eqref{may5-13-20} requires that $\gamma$ is sufficiently small.
The condition \eqref{may3-13-6} is satisfied automatically,
provided that the coefficients in \eqref{may5-13-11} obey 
eq.~\eqref{apr19-13-10}. Note that since $b (0) > 0$ and $\kappa (0) =0$,
pressure \eqref{apr19-13-7} is {\it positive}
at small $\Phi$. Away from the large sphere with $\Phi = \Phi_0$, the space
{\it contracts}.

To see explicitly that all above conditions can be satisfied,
let us choose
\[
b (\Phi) = u + v\Phi^2 \; , \;\;\; \kappa (\Phi) = 0 \;  , \;\;\;
\beta (\Phi) = w \Phi  \;  , \;\;\; c (\Phi) > \beta (\Phi)
 \;  , \;\;\; \gamma (\Phi) = 0 
\]
with constant $u>0$, $w>0$, $v<0$
and $w \Phi_0^2 \gg u$, while $a(\Phi)$ is given by
eq.~\eqref{apr19-13-10}. Then the only non-trivial constraints are
\eqref{may5-13-12} and \eqref{may5-13-13}. These are satisfied by choosing
\[
   -4 w < v < -w \; .
\]
Thus, there is indeed a choice of
 $F(Y;\varphi)$, $K(Y;\varphi)$ such that the entire set up
is not pathological everywhere in space (including the wall region), 
at least in the quasi-homogeneous case.

\subsection{Sample scenario}
\label{improved-scenario}

\begin{figure}[!htb]
\centerline{\includegraphics[width=0.5\textwidth,angle=-90]{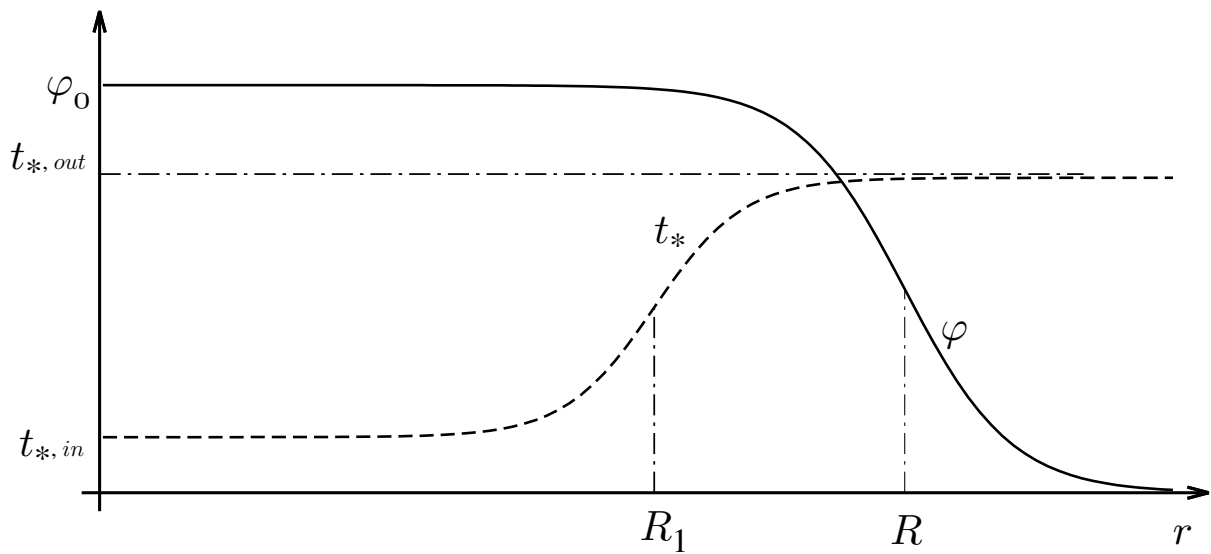}%
}
\caption{The set up. Dashed and solid lines show $t_* (r)$ and $\varphi (r)$,
respectively. The behaviour of the
function $\Phi (r) = \Phi (\varphi (r))$ is similar 
to that of $\varphi (r)$.
\label{fig1}
}
\end{figure}

Let us now sketch a concrete scenario for creating a universe. 
Let us assume that
the field $\varphi$ is a usual scalar field which has two
vacua, $\varphi = 0$ and $\varphi = \varphi_0$. We prepare
a spherical configuraion of this field with  $\varphi = \varphi_0$
inside a sphere of large enough radius $R$ and  $\varphi = 0$
outside this sphere, see Fig.~\ref{fig1}. 
We assume for definiteness (although this assumption can be relaxed)
that there is a source for the field
$\varphi$ that keeps this configuration static.
Let $L \ll R$ be the thickness of the wall separating
the two vacua; $L$ is also kept time-independent by the source.
We require that the mass of this ball is small enough, so that
$R \gg R_s$, where $R_s$ is the Schwarzschild radius.
The mass is of order $\mu^4 R^2 L$, where $\mu$ is the mass scale 
characteristic of the field $\varphi$. Hence, the latter requirement
reads $\mu^4 R L \ll M_{Pl}^2$. For small enough $\mu$ both $R$ and $L$
can be large.

Let the function $\Phi (\varphi)$ of Sec. \ref{sec:improved1}
be such that $\Phi (0) = 0$, $\Phi (\varphi_0) = \Phi_0$ and
$\Phi^\prime (\varphi_0) = 0$. The latter property ensures that
coupling of $\pi$ to $\varphi$ does not move $\varphi$ out of the
vacuum $\varphi_0$ inside the large sphere, whatever $\pi$ does 
there\footnote{We implicitly neglect kinetic mixing between
$\pi$ and $\varphi$. It can be made small by considering the function
$\Phi$ which depends on $\lambda \varphi$, where $\lambda$ is a small
parameter.}. 

We prepare the initial configuration of $\pi$ at $t=0$ in such a way
that it initially evolves as
\be
\e^{\pi} = \frac{1}{\sqrt{\Phi_0}t_* (r) - 
\sqrt{\Phi(r)}t} \; ,
\label{may9-13-1}
\ee
where we allow the parameter $t_*$ in \eqref{may4-13-1} to vary in space,
and choose a convenient parametrization. We choose $t_* (r)=t_{*, in}$ 
inside a somewhat smaller sphere of radius $R_1 < R$
(but $R_1 \sim R$) and 
$t_*(r)=t_{*, out} \gg t_{*, in}$ at $r > R_1$ (hereafter
subscripts {\it in} and {\it out}
refer to the regions $r < R_1$ and $r> R_1$, respectively),
as shown in Fig.~\ref{fig1},  with the transition region
of, say, the same thickness $L$.
We take $t_{*, out} \ll L$, then the
characteristic time scales are smaller than the smallest length scale $L$
inherent in the set up,
so the spatial devivatives of $\pi$ are negligible compared to 
the time derivatives. This ensures that
the field $\pi$ is in the quasi-homogeneous regime.
As $r \to \infty$, we have $\Phi(r) \to 0$ and $t_* \to \mbox{const}$, so
 the field $\pi$ tends to the
Minkowski vacuum $\pi = \mbox{const}$.

At the initial stage of evolution, 
pressure inside the sphere of radius $R_1$ is
\[
p_{in} = - \frac{M^4}{\Phi_0^2 (t_{*, in} - t)^4} \; ,
\]
where $M$ is the mass scale characteristic of the field $\pi$.
We require that 
$|p_{in}| R^3/M_{Pl}^2 \ll R $,
then the gravitational potentials are small everywhere,
and gravity is initially
in the linear regime. Thus, we impose a constraint
\be
    \frac{M^4 R^2}{\Phi_0^2 t_{* , in}^4} \ll M_{Pl}^2 \; ,
\label{may7-13-1}
\ee
which is consistent with the above conditions for
$M \ll M_{Pl}$ and $\Phi_0 \gtrsim M^2$. 

At least at the initial stage of the evolution,
the field $\pi(r,t)$ is in the quasi-homogeneous regime and 
evolves according to \eqref{may9-13-1}. The metric is also
quasi-homogeneous,
\[
ds^2 = dt^2 - a^2 (r,t) \left( dr^2 + r^2 d\Omega^2 \right)
\; .
\]
We integrate the equation
$\dot{H} = - 4\pi G p$ to find that
soon after the evolution begins, the Hubble parameter
inside the sphere of radius $R_1$ is
\be
H_{in} = \frac{4 \pi M^4}{3 M_{Pl}^2 \Phi_0^2 (t_{* , in} - t)^3} \; .
\label{may9-13-2}
\ee
In view of \eqref{may7-13-1} and $t_{* ,in} \ll R$, the Hubble
length scale is  large for some time, $H^{-1} \gg R$.
This is true also at $r > R_1$, so
there are no anti-trapped surfaces initially. 

As $t$ approaches $t_{*, in}$, pressure becomes large at $r< R_1$
and the Hubble length shrinks there to $R_1 \sim R$. The 
anti-trapped surfaces get formed inside the sphere of radius $R_1$,
a new universe gets created and enters the Genesis regime there. 
This occurs when $H_{in} \sim R^{-1}$, i.e., at time $t_1$ such that
\[
(t_{* , in} - t_1) \sim \l \frac{M^4 R}{M_{Pl}^2 \Phi_0^2 } \r^{1/3}\; .
\]
Note that at that time the energy density 
$\rho_{in} \sim M_{Pl}^2 H_{in}^2$ is still relatively small,
\[
\frac{\rho_{in}}{|p_{in}|} \sim \l \frac{M^4}{\Phi_0^2 R^2 M_{Pl}^2} \r^{1/3} 
\ll 1 \; .
\]
This implies that at time $t_1$, space-time is locally nearly
Minkowskian. Another manifestation of this fact is that the scale
factor is close to 1:
\[
a_{in}(t_1) = 1 + \frac{2\pi M^4}{3 M_{Pl}^2 \Phi_0^2 (t_{* , in} - t_1)^2}
\]
where the correction to 1 is of order $\rho_{in}/|p_{in}|$. Hence,
our approximate solution \eqref{may9-13-1}, \eqref{may9-13-2} is legitimate.

Since $t_{* , out} \gg t_{* , in}$, the field $e^{\pi}$ at time $t_1$
is still small
at $r>R_1$, and the Hubble length scale exceeds $R$ there.
Gravity is still weak at $r>R_1$, so it is consistent to assume that
the configuration of $\varphi$ is not modified by that time.
Note also that
a black hole is not formed by then either.

At somewhat later times, the geometry of hypersurfaces
$t=\mbox{const}$ is that of a semi-closed world:
at some distance from the origin the area of the sphere
$r = \mbox{const}$ decreases as $r$ increases, $\d (ar)/\d r < 0$.
This regime begins (i.e., a neck gets formed)
when there appears a solution to
\[
\frac{\d a}{\d r} \cdot r + a \approx
- \frac{4\pi M^4}{3 M_{Pl}^2 \Phi_0^2 [t_*(r) - t]^3} \frac{\d t_*}{\d r}\cdot r
+ 1 = 0 \; .
\]
Clearly, this happens at
a place where $t_* (r) \simeq t_{* , in}$, but
$\d t_*/\d r$ starts 
to deviate
from zero. The neck gets formed at time $t_2$ such that
\[
\frac{M^4}{M_{Pl}^2 \Phi_0^2 (t_{* , in} - t_2)^3} \frac{ t_{*, out} R}{L} 
\sim 1 \; .
\]
Since we take $t_{* , out} \ll L$, we have $t_2 > t_1$ indeed. Nevertheless,
it is straightforward to arrange parameters in such a way that
our approximate solution \eqref{may9-13-1}, \eqref{may9-13-2} is legitimate
at time $t_2$ as well.

This completes the discussion of the initial stage of the creation of
a new universe. To make the scenario complete, one would
specify the way to design the configuration of the field $\varphi$
and keep it static (or consider an evolving field $\varphi$ instead). 
Also, one would like to trace the dynamics of the system to longer
times and see what geometry develops towards the end of the Genesis
epoch occuring at $r<R_1$.
Since the background we have studied is healthy, one does not
expect surprizes from this complete analysis.

\section{Discussion}
\label{concl}

Because of the obstruction we encountered in Sec.~\ref{sec:obstruction},
it is rather unlikely that simple, scale-invariant
Galileon-type theories can be
employed to create a universe in the laboratory. We had to make the
model a lot more complicated, to the extent that the whole scenario 
may appear
completely unrealistic. While the particular model we considered
in Sec.~\ref{improved-all} is indeed not very appealing, 
we think that the overall situation is not absolutely hopeless.
First, one can think of a possibility that a model designed on
paper can be implemented in the laboratory, even though this certainly
sounds as fiction. Barring this possibility, let us make the second point.
If there is anything like Galileon in Nature, and if the Universe
experienced anything like the 
Genesis epoch, there {\it must be} a smooth and
consistent interpolation between the Genesis regime and 
Minkowski vacuum, albeit in the course of the cosmological evolution
rather than in the radial direction in space as we need. 
It is not inconceivable
that one may be able to use the mechanism making this interpolation
healthy in cosmology
for the purpose of creating a universe in the laboratory.

\vspace{0.3cm}

The author is indebted to S.~Demidov, D.~Levkov, M.~Libanov, 
I.~Tkachev and M.~Voloshin
for helpful discussions
and S.~Deser,   Y.-S.~Piao and  A.~Vikman for useful
correspondence. This work has been supported in part
by RFBR grant 12-02-0653.


\begin{thebibliography}{99}

\bibitem{Berezin:1984vy} 
  V.~A.~Berezin, V.~A.~Kuzmin and I.~I.~Tkachev,
  ``Dynamics Of Inflating Bubbles In The Early Universe,''
  In {\it Proc. 3d Seminar on Quantum Gravity, Moscow, 1984;
World Scientific, Singapore, 1985}, 
605-622;\\
V.~A.~Berezin, V.~A.~Kuzmin and I.~I.~Tkachev,
  JETP Lett.\  {\bf 41}, 547 (1985)
  [Pisma Zh.\ Eksp.\ Teor.\ Fiz.\  {\bf 41}, 446 (1985)].

\bibitem{Farhi:1986ty} 
  E.~Farhi and A.~H.~Guth,
  Phys.\ Lett.\ B {\bf 183}, 149 (1987).

\bibitem{Berezin:1987ep} 
  V.~A.~Berezin, V.~A.~Kuzmin and I.~I.~Tkachev,
  Sov.\ Phys.\ JETP {\bf 66}, 654 (1987)
  [Zh.\ Eksp.\ Teor.\ Fiz.\  {\bf 93}, 1159 (1987)].

\bibitem{Penrose}
R. Penrose,
Phys.\ Rev.\ Lett.\ {\bf 14}, 57 (1965)

\bibitem{Borde:1998wa} 
  A.~Borde, M.~Trodden and T.~Vachaspati,
  Phys.\ Rev.\ D {\bf 59}, 043513 (1999)
  [gr-qc/9808069].

\bibitem{Sakai:2006fg} 
  N.~Sakai, K.~-i.~Nakao, H.~Ishihara and M.~Kobayashi,
  Phys.\ Rev.\ D {\bf 74}, 024026 (2006)
  [gr-qc/0602084].

\bibitem{Berezin:1987ea} 
  V.~A.~Berezin, V.~A.~Kuzmin and I.~I.~Tkachev,
  Phys.\ Lett.\ B {\bf 207}, 397 (1988).

\bibitem{Farhi:1989yr} 
  E.~Farhi, A.~H.~Guth and J.~Guven,
  Nucl.\ Phys.\ B {\bf 339}, 417 (1990).

\bibitem{Fischler:1989se} 
  W.~Fischler, D.~Morgan and J.~Polchinski,
  Phys.\ Rev.\ D {\bf 41}, 2638 (1990); \\
%
%
  Phys.\ Rev.\ D {\bf 42}, 4042 (1990).



\bibitem{Linde:1991sk} 
  A.~D.~Linde,
  Nucl.\ Phys.\ B {\bf 372}, 421 (1992)
  [hep-th/9110037].

\bibitem{Garriga:1997ef} 
  J.~Garriga and A.~Vilenkin,
  Phys.\ Rev.\ D {\bf 57}, 2230 (1998)
  [astro-ph/9707292].

\bibitem{Garriga:1999hf} 
  J.~Garriga, V.~F.~Mukhanov, K.~D.~Olum and A.~Vilenkin,
  Int.\ J.\ Theor.\ Phys.\  {\bf 39}, 1887 (2000)
  [astro-ph/9909143].

\bibitem{Dutta:2005gt} 
  S.~Dutta and T.~Vachaspati,
  Phys.\ Rev.\ D {\bf 71}, 083507 (2005)
  [astro-ph/0501396].

\bibitem{Aguirre:2005nt} 
  A.~Aguirre and M.~C.~Johnson,
  Phys.\ Rev.\ D {\bf 73}, 123529 (2006)
  [gr-qc/0512034].

\bibitem{Lee:2006vka} 
  W.~Lee, B.~-H.~Lee, C.~H.~Lee and C.~Park,
  Phys.\ Rev.\ D {\bf 74}, 123520 (2006)
  [hep-th/0604064].

\bibitem{Piao:2007cj} 
  Y.~-S.~Piao,
  Nucl.\ Phys.\ B {\bf 803}, 194 (2008)
  [arXiv:0712.4184 [gr-qc]].

\bibitem{Frolov:1988vj} 
 V.~P.~Frolov, M.~A.~Markov and V.~F.~Mukhanov,
  Phys.\ Lett.\ B {\bf 216}, 272 (1989);\\
  Phys.\ Rev.\ D {\bf 41}, 383 (1990).

\bibitem{Guendelman:2010pr} 
  E.~I.~Guendelman,
  Int.\ J.\ Mod.\ Phys.\ D {\bf 19}, 1357 (2010)
  [arXiv:1003.3975 [gr-qc]].

\bibitem{Lukash:2013ts} 
  V.~N.~Lukash and V.~N.~Strokov,
  Int.\ J.\ Mod.\ Phys.\ A {\bf 28}, 1350007 (2013)
  [arXiv:1301.5544 [gr-qc]].

\bibitem{Mukhanov:1991zn} 
  V.~F.~Mukhanov and R.~H.~Brandenberger,
  Phys.\ Rev.\ Lett.\  {\bf 68}, 1969 (1992).

\bibitem{Brandenberger:1993ef} 
  R.~H.~Brandenberger, V.~F.~Mukhanov and A.~Sornborger,
  Phys.\ Rev.\ D {\bf 48}, 1629 (1993)
  [gr-qc/9303001].

\bibitem{Trodden:1993dm} 
  M.~Trodden, V.~F.~Mukhanov and R.~H.~Brandenberger,
  Phys.\ Lett.\ B {\bf 316}, 483 (1993)
  [hep-th/9305111].

\bibitem{Lee:2007dh} 
  B.~-H.~Lee, C.~H.~Lee, W.~Lee, S.~Nam and C.~Park,
  Phys.\ Rev.\ D {\bf 77}, 063502 (2008)
  [arXiv:0710.4599 [hep-th]].

\bibitem{Guendelman:2008ys} 
  E.~I.~Guendelman and N.~Sakai,
  Phys.\ Rev.\ D {\bf 77}, 125002 (2008)
  [Erratum-ibid.\ D {\bf 80}, 049901 (2009)]
  [arXiv:0803.0268 [gr-qc]].


\bibitem{Yeom:2009mn} 
  D.~-h.~Yeom,
  arXiv:0912.0068 [gr-qc].

\bibitem{Hwang:2010gc} 
  D.~-i.~Hwang and D.~-h.~Yeom,
  Class.\ Quant.\ Grav.\  {\bf 28}, 155003 (2011)
  [arXiv:1010.3834 [gr-qc]].

\bibitem{Senatore:2004rj} 
  L.~Senatore,
  Phys.\ Rev.\ D {\bf 71}, 043512 (2005)
  [astro-ph/0406187].


\bibitem{VR-vector}
  V.~A.~Rubakov,
  Theor.\ Math.\ Phys.\  {\bf 149}, 1651 (2006)
  [Teor.\ Mat.\ Fiz.\  {\bf 149}, 409 (2006)]
  [hep-th/0604153];\\
M.~Libanov, V.~Rubakov, E.~Papantonopoulos, M.~Sami and S.~Tsujikawa,
  JCAP {\bf 0708}, 010 (2007)
  [arXiv:0704.1848 [hep-th]].

\bibitem{Creminelli:2006xe} 
  P.~Creminelli, M.~A.~Luty, A.~Nicolis and L.~Senatore,
  JHEP {\bf 0612}, 080 (2006)
  [hep-th/0606090].

\bibitem{Nicolis:2009qm} 
  A.~Nicolis, R.~Rattazzi and E.~Trincherini,
  JHEP {\bf 1005}, 095 (2010)
  [Erratum-ibid.\  {\bf 1111}, 128 (2011)]
  [arXiv:0912.4258 [hep-th]].

\bibitem{Genesis1}
  P.~Creminelli, A.~Nicolis and E.~Trincherini,
  JCAP {\bf 1011}, 021 (2010)
  [arXiv:1007.0027 [hep-th]].

\bibitem{Genesis2}
  P.~Creminelli, K.~Hinterbichler, J.~Khoury, A.~Nicolis and E.~Trincherini,
  JHEP {\bf 1302}, 006 (2013)
  [arXiv:1209.3768 [hep-th]].


\bibitem{Nicolis:2008in} 
  A.~Nicolis, R.~Rattazzi and E.~Trincherini,
  Phys.\ Rev.\ D {\bf 79}, 064036 (2009)
  [arXiv:0811.2197 [hep-th]].

\bibitem{deRham:2010eu} 
  C.~de Rham and A.~J.~Tolley,
  JCAP {\bf 1005}, 015 (2010)
  [arXiv:1003.5917 [hep-th]].

\bibitem{Deffayet:2010qz} 
  C.~Deffayet, O.~Pujolas, I.~Sawicki and A.~Vikman,
  JCAP {\bf 1010}, 026 (2010)
  [arXiv:1008.0048 [hep-th]].

\bibitem{Kobayashi:2010cm} 
  T.~Kobayashi, M.~Yamaguchi and J.~'i.~Yokoyama,
  Phys.\ Rev.\ Lett.\  {\bf 105}, 231302 (2010)
  [arXiv:1008.0603 [hep-th]].

\bibitem{Goon-prl}
G.~Goon, K.~Hinterbichler and M.~Trodden,
  Phys.\ Rev.\ Lett.\  {\bf 106}, 231102 (2011)
  [arXiv:1103.6029 [hep-th]].

\bibitem{Goon:2011qf} 
  G.~Goon, K.~Hinterbichler and M.~Trodden,
  JCAP {\bf 1107}, 017 (2011)
  [arXiv:1103.5745 [hep-th]].

\bibitem{Deffayet:2010zh} 
  C.~Deffayet, S.~Deser and G.~Esposito-Farese,
  Phys.\ Rev.\ D {\bf 82}, 061501 (2010)
  [arXiv:1007.5278 [gr-qc]].

\bibitem{Kamada:2010qe} 
  K.~Kamada, T.~Kobayashi, M.~Yamaguchi and J.~'i.~Yokoyama,
  Phys.\ Rev.\ D {\bf 83}, 083515 (2011)
  [arXiv:1012.4238 [astro-ph.CO]];\\
%
  T.~Kobayashi, M.~Yamaguchi and J.~'i.~Yokoyama,
  Prog.\ Theor.\ Phys.\  {\bf 126}, 511  (2011)
  [arXiv:1105.5723 [hep-th]].

\bibitem{Pujolas:2011he} 
  O.~Pujolas, I.~Sawicki and A.~Vikman,
  JHEP {\bf 1111}, 156 (2011)
  [arXiv:1103.5360 [hep-th]];\\
%
%
  D.~A.~Easson, I.~Sawicki and A.~Vikman,
  JCAP {\bf 1111}, 021 (2011)
  [arXiv:1109.1047 [hep-th]].


\bibitem{Adams:2006sv} 
  A.~Adams, N.~Arkani-Hamed, S.~Dubovsky, A.~Nicolis and R.~Rattazzi,
  JHEP {\bf 0610}, 014 (2006)
  [hep-th/0602178].

\bibitem{Easson:2013bda} 
  D.~A.~Easson, I.~Sawicki and A.~Vikman,
  ``When Matter Matters,''
  arXiv:1304.3903 [hep-th].


\end{thebibliography}
\end{document}